 \documentclass[preprint,aps]{revtex4}

\usepackage{graphicx}
\usepackage{bm}

\begin{document}
%%\title{Nodal Superconductivity in Nematic Phase of Bulk FeSe Superconductor}
%%\title{Direct Evidence on Orbital Origin of Highly Anisotropic Superconducting Gap in Nematic Phase of FeSe Superconductor}
%%\title{Direct Determination on Orbital Origin of Highly Anisotropic Superconducting Gap in Nematic Phase of FeSe Superconductor}
%%\title{Orbital Origin of Highly Anisotropic Superconducting Gap in Nematic Phase of FeSe Superconductor}
\title{Orbital Origin of Extremely Anisotropic Superconducting Gap in Nematic Phase of FeSe Superconductor}

\author{Defa Liu$^{1,3\sharp}$, Cong Li$^{1,2\sharp}$, Jianwei Huang$^{1,2\sharp}$, Bin Lei$^{4}$, Le Wang$^{1,2}$, Xianxin Wu$^{5}$, Bing Shen$^{1,2}$, Qiang Gao$^{1,2}$, Yuxiao Zhang$^{1}$, Xu Liu$^{1}$, Yong Hu$^{1,2}$, Yu Xu$^{1,2}$, Aiji Liang$^{1}$, Jing Liu$^{1,2}$, Ping Ai$^{1,2}$, Lin Zhao$^{1}$, Shaolong He$^{1}$, Li Yu$^{1}$, Guodong Liu$^{1}$, Yiyuan Mao$^{1,2}$, Xiaoli Dong$^{1}$, Xiaowen Jia$^{6}$, Fengfeng Zhang$^{7}$, Shenjin Zhang$^{7}$, Feng Yang$^{7}$, Zhimin Wang$^{7}$, Qinjun Peng$^{7}$, Youguo Shi$^{1}$, Jiangping Hu$^{1,2,8}$, Tao Xiang$^{1,2,8}$, Xianhui Chen$^{4}$, Zuyan Xu$^{7}$, Chuangtian Chen$^{7}$ and X. J. Zhou$^{1,2,8,*}$
}

\affiliation{
\\$^{1}$Beijing National Laboratory for Condensed Matter Physics, Institute of Physics, Chinese Academy of Sciences, Beijing 100190, China
\\$^{2}$University of Chinese Academy of Sciences, Beijing 100049, China
\\$^{3}$Max Planck Institute of Microstructure Physics, Weinberg 2, Halle 06120, Germany
\\$^{4}$Hefei National Laboratory for Physical Science at Microscale and Department of Physics, University of Science and Technology of China, Hefei 230026, China
\\$^{5}$Insitute of Theoretical Physics and Astrophysics, Julius-Maximilaians University of Wurzburg, Am Hubland, Wurzburg D-97074, Germany
\\$^{6}$Military Transportation University, Tianjin 300161, China
\\$^{7}$Technical Institute of Physics and Chemistry, Chinese Academy of Sciences, Beijing 100190, China
\\$^{8}$Collaborative Innovation Center of Quantum Matter, Beijing 100871, China
\\$^{\sharp}$These people contributed equally to the present work.
\\$^{*}$Corresponding authors: XJZhou@iphy.ac.cn
}

\date{Feb. 8, 2018}


\maketitle
%1 多轨道体系；2确定哪些轨道对超导的决定性作用；3多轨道产生新的物理， 轨道选择性(莫特选择性\YIMin_NC），轨道有序(orbital order, orbital 涨落 s++；40
%Abstract
{\bf The iron-based superconductors are characterized by multiple-orbital physics where all the five Fe $\boldsymbol{3d}$ orbitals get involved.
The multiple-orbital nature gives rise to various novel phenomena like orbital-selective Mott transition, nematicity and orbital fluctuation that provide a new route for realizing superconductivity.
%It is important to identity dominant orbital ingredients that dictate superconductivity in the iron-based superconductors.
The complexity of multiple-orbital also asks to disentangle the relationship between orbital, spin and nematicity, and to identify dominant orbital ingredients that dictate superconductivity.
The bulk FeSe superconductor provides an ideal platform to address these issues because of its simple crystal structure and unique coexistence of superconductivity and nematicity.
However, the orbital nature of the low energy electronic excitations and its relation to the superconducting gap remain controversial.
Here we report direct observation of highly anisotropic Fermi surface and extremely anisotropic superconducting gap in the nematic state of FeSe superconductor by high resolution laser-based angle-resolved photoemission measurements. We find that the low energy excitations of the entire hole pocket at the Brillouin zone center are dominated by the single $\boldsymbol{d_{xz}}$ orbital. The superconducting gap exhibits an anti-correlation relation with the $\boldsymbol{d_{xz}}$ spectral weight near the Fermi level, i.e., the gap size minimum (maximum) corresponds to the maximum (minimum) of the $\boldsymbol{d_{xz}}$ spectral weight along the Fermi surface. These observations provide new insights in understanding the orbital origin of the extremely anisotropic superconducting gap in FeSe superconductor and the relation between nematicity and superconductivity in the iron-based superconductors.}

In the iron-based superconductors, all the five Fe $3d$ orbitals ($d_{xz}$, $d_{yz}$, $d_{xy}$, $d_{x^2-y^2}$ and $d_{z^2}$) are involved in the low energy electronic excitations\cite{Lebegue_PRB,Paglione_NP}. The multiple-orbital character provides a new degree of freedom which, when combined with charge and spin, brings new phenomena like orbital-selective Mott transition\cite{MYi_PRL,Rong_PRL}, orbital ordering\cite{Frank_PRB,CCLee_PRL} and nematicity\cite{Fernandes_NP}. Orbital fluctuation may provide a new channel for realizing superconductivity\cite{Saito_PRB,Kontani_PRL}. On the other hand, such multiple-orbital nature also brings complexity in finding the key ingredients of superconductivity in the iron-based superconductors. FeSe is unique in the iron-based superconductors because it has the simplest crystal structure\cite{Hsu_PNAS}. It shows a nematic transition at $\sim$90 K without being accompanied by a magnetic transition\cite{McQueen_PRB}. In particular, FeSe superconductor provides an ideal case for studying the relationship between nematicity and superconductivity\cite{Glasbrenner_NP2015,FWang_NP2015,RYu_PRL,XXWu_arxiv} because superconductivity occurs in the nematic state. However the experimental results are controversial regarding the superconducting gap of FeSe on whether it is nodeless\cite{Dong_PRB,Hope_PRL,Lin_PRB,Jiao_arxiv,Hafiez_PRB,Khasanov_PRL,Sprau_arxiv} or nodal\cite{Song_Science,Kasahara_PNAS,THashimoto_NC}.
It is also under debate on the orbital nature of the low energy excitations that dictates superconductivity\cite{Sprau_arxiv,Suzuki_PRB,Kreisel_PRB}.
Direct determination of the correspondence between the orbital nature of the low energy electronic states and the superconducting gap is crucial to understand the superconductivity mechanism of the iron-based superconductors.

In this paper, we performed high resolution laser-based angle-resolved photoemission (ARPES) measurements on the electronic structure and superconducting gap of bulk FeSe superconductor (T$_{c}$=8.0 K) in the nematic state. Highly anisotropic Fermi surface around the Brillouin zone (BZ) center is observed with the aspect ratio of $\sim$3 between the long axis (along $k_{y}$) and short axis (along $k_{x}$). The superconducting gap along the Fermi pocket is extremely anisotropic, varying between $\sim$3 meV along the short axis of Fermi surface to zero along the long axis within our experimental precision ($\pm$0.2 meV). Detailed band structure analysis, combined with band structure calculations, indicates that the Fermi surface is dominated by a single $d_{xz}$ orbital. Moreover, we find that the superconducting gap size shows an anti-correlation with the $d_{xz}$ spectral weight near the Fermi level; the gap minimum (maximum) corresponds to the $d_{xz}$ spectral intensity maximum (minimum) along the Fermi surface. These observations provide key insights on the orbital origin of the anisotropic electronic structure and superconducting gap in FeSe and the interplay between nematicity and superconductivity in the iron-based superconductors.

The electronic structure and superconducting gap of FeSe superconductor (T$_{c}$=8.0 K) (see Methods and Fig. S1 for sample details) were measured by high resolution laser-based angle-resolved photoemission system based on the time-of-flight electron energy analyzer (see Methods and Fig. S2 for experimental details). This new ARPES system has an advantage of covering two-dimensional momentum space at the same time with high energy resolution ($\sim$1 meV) and momentum resolution. It is also equipped with an ultra-low temperature cryostat which can cool down the sample to 1.6 K. The laser polarization can be tuned to identify the orbital character of the observed band structure by taking the advantage of the photoemission matrix element effects\cite{Damascelli_RMP}. This system is particularly suitable for bulk FeSe because of its low T$_{c}$, tiny Fermi pockets and very small superconducting gap.

%Figure1
The FeSe sample we measured has a structural phase transition T$_{s}$ around 90 K (Fig. S1b). It assumes a tetragonal crystal structure above T$_{s}$ and becomes orthorhombic below T$_{s}$ where the distance between adjacent Fe is slightly different (Fig. 1a, where $a_{Fe}(b_{Fe})$ is defined as the long $x$ (short $y$) axis). Dramatic anisotropy of physical properties between $a_{Fe}$ and $b_{Fe}$ axes occurs below the nematic transition temperature\cite{Tanatar_PRL,AEBohmer_PRL,Baek_NM,Boehmer} although the lattice constant difference between $a_{Fe}$ and $b_{Fe}$ is only $\sim$0.5$\%$\cite{Margadonna_CC}. Fig. 1b shows the measured Fermi surface of FeSe around the BZ center which occupies a small portion of the entire BZ. Our laser-based ARPES system made it possible to cover the whole Fermi pocket around the $\Gamma$ point simultaneously under the same condition with very dense momentum points. Fig. 1c shows constant energy contours of FeSe at different binding energies measured at 1.6 K. The measured Fermi surface (leftmost panel of Fig. 1c) is highly anisotropic: the Fermi momentum along the $k_{x}$ direction is $\sim$0.036$\pi/a$ while it is $\sim$0.11$\pi/a$ along the $k_{y}$ direction resulting in an elongated ellipse with a high aspect ratio of $\sim$3. This is the Fermi surface that exhibits the strongest anisotropy observed among all the iron-based superconductors\cite{DFLiu_CPL}. For the photon energy we used (6.994 eV), the measured Fermi surface corresponds to a $k_{z}$ close to zero\cite{Watson_PRB}. With increasing binding energy, the constant energy contours increase in area (Fig. 1c), consistent with the Fermi pocket being hole-like. In the meantime, the spectral weight is gradually concentrated onto the two vertex areas along the long axis while it gets strongly suppressed in the central region. We note that for the FeSe single crystal sample we measured here it has nearly pure single domain even though we did not detwin the sample in advance. This may be due to accidental internal stress exerted in the sample during the preparation process because in most cases we can see signals from coexisting two domains (Fig. S2). In the FeSe sample we measured (Fig. 1), the signal from another domain is negligible even under different light polarizations. The single domain FeSe sample provided us an ideal opportunity to study its intrinsic electronic structure and superconducting gap as we will present below.

Figure 1d shows photoemission spectra (energy distribution curves, EDCs) of FeSe measured at two typical momentum points along the Fermi surface (A and B points as marked in Fig. 1c) at different temperatures. The corresponding symmetrized EDCs are shown in Fig. 1e. When the sample cools down from the normal state to the superconducting state, for both momentum points A and B, sharp superconducting coherence peaks develop which get stronger with decreasing temperature (Fig. 1d). The sharp coherence peak with a width of $\sim$6 meV is observed at 1.6 K. The EDC symmetrization is a standard procedure to remove the Fermi-Dirac function in order to extract the superconducting gap\cite{Norman_PRB}. The gap size can be determined by the distance between the two peaks in the symmetrized EDCs which can be fitted by a phenomenological gap formula\cite{Norman_PRB}. For point A, the symmetrized EDCs always show a peak at the Fermi level (left panel in Fig. 1e), indicating no detectable gap opening in the superconducting state within our experimental resolution. On the other hand, for point B, the symmetrized EDCs show clear two peaks in the superconducting state (right panel in Fig. 1e) indicative of superconducting gap opening. The extracted superconducting gap at different temperatures (Fig. 1f) follows a BCS gap form; it becomes zero at and above the superconducting transition temperature of 8.0 K.

%Figure2
Figure 2 shows detailed momentum dependence of the superconducting gap for FeSe measured at 1.6 K. The symmetrized EDCs at some representative Fermi momenta are shown in Fig. 2a. Fig. 2b is a false color image which represents the same data as in Fig. 2a. The location of the Fermi momenta is defined by the Fermi surface angle $\theta$ as indicated in Fig. 2c. As seen directly from Fig. 2a and 2b, the superconducting gap shows a maximum near $\theta$=0$^\circ$ and 180$^\circ$ and becomes rather small, basically zero within our experimental precision ($\pm$0.2 meV) near $\theta$=90$^\circ$ and 270$^\circ$. This is consistent with the recent result that the superconducting gap approaches zero near $\theta$=90$^\circ$ region although their gap measurement covers a small portion of the Fermi surface in detwinned FeSe\cite{THashimoto_NC}. The symmetrized EDCs are fitted by the phenomenological gap formula, as shown in Fig. 2a, and the extracted superconducting gap is plotted in Fig. 2d as function of the Fermi surface angle. The simultaneous two-dimensional momentum coverage of our laser ARPES system makes it possible to have very dense data points of superconducting gap along the entire Fermi surface measured with super-high resolution. This allows us to quantitatively examine on various possible gap functions. The two-fold symmetry of the superconducting gap has excluded the possibility of pure $s$-wave which would have four-fold symmetry. The two-fold symmetry, together with two near-zero points, is rather reminiscent of a $p$-wave gap form. The measured gap is fitted by the $p$-wave form $\Delta_{p}$=$|\Delta_{0}cos(\theta)|$ and the fitted curve is shown as a green line in Fig. 2d. An alternative gap form is $s$+$d$ type which can also assume two-fold symmetry\cite{JKang_PRL,Xu_PRL}. In Fig. 2d, we also fitted the measured data with two types of $s$+$d$ form. We first tried $\Delta_{s,d}$=$|\Delta_{0}$+$\Delta_{1}cos(2\theta)|$ which contains a simple $s$-wave form $\Delta_{0}$ and a $d$-wave form $\Delta_{1}cos(2\theta)$. The fitted curve (blue line in Fig. 2d) exhibits an obvious deviation from the measured data, particularly near the minimal and maximal gap regions. Then we tried $\Delta_{es,d}$=$|\Delta_{0}$+$\Delta_{1}cos(2\theta)$+$\Delta_{2}cos(4\theta)|$ which contains an anisotropic $s$-wave $\Delta_{0}$+$\Delta_{2}cos(4\theta)$ and a $d$-wave form $\Delta_{1}cos(2\theta)$. The fitted curve is marked as a red line in Fig. 2d. We note that within our experimental precision ($\pm$0.2 meV), we can not differentiate between the cases of zero node, two nodes and four nodes along the Fermi surface for the $es$+$d$ gap form (Fig. S6). The observation of two nodes on the Fermi surface in terms of $es$+$d$ gap form is quite accidental because three fitting parameters are needed and a constraint has to be imposed between these parameters, i.e., ($\Delta_{0}$+$\Delta_{2}$)-$\Delta_{1}$=0 (Fig. S6e). Our precise measurement on the momentum dependence of the superconducting gap puts a strong constraint on the possible gap form in FeSe superconductor.

%Figure3
Figure 3 shows detailed momentum dependence of the band structure for FeSe measured at 1.6 K, along the momentum cuts crossing the BZ center with different Fermi surface angles defined as $\theta$ in Fig. 3a. Two main bands are observed for all the momentum cuts varying from horizontal ($\theta$=0$^\circ$) to vertical ($\theta$=90$^\circ$) directions, labeled as $\alpha$ and $\beta$ bands shown in the leftmost and rightmost panels of Fig. 3b. When the momentum cuts vary from vertical ($\theta$=90$^\circ$) to horizontal ($\theta$=0$^\circ$) directions, the Fermi momentum of the $\alpha$ band significantly decreases and the band gets steeper. The top of this $\alpha$ band lies at $\sim$6.7 meV above the Fermi level, as estimated by fitting the hole-like band with a parabola (red curve in the leftmost panel in Fig. 3b). The top of the $\beta$ band stays at $\sim$20 meV below the Fermi level (dashed green line in the leftmost panel in Fig. 3b). This band gets flatter when the momentum cuts change from vertical ($\theta$=90$^\circ$) to horizontal ($\theta$=0$^\circ$) directions. Based on the analysis of the photoemission matrix elements (see Fig. S2), the $\alpha$ and $\beta$ bands can be attributed to the $d_{xz}$ and $d_{yz}$ orbitals, respectively. This assignment is further supported by our band structure calculations (Fig. 3c and 3d) which capture the momentum dependence of the two bands well (see Methods and Supplementary materials). Above the nematic transition, $d_{xz}$ and $d_{yz}$ bands are split because of the spin orbital coupling (Fig. S3a and S3b)\cite{Watson_PRB}. In the nematic state, the $d_{xz}$ band is pushed up while the $d_{yz}$ band is pushed down below the Fermi level (Fig. S3c and S3d)\cite{Suzuki_PRB}, resulting in only one hole-like Fermi pocket around $\Gamma$ which is dominated by the $d_{xz}$ orbital. In particular, the electronic states at the Fermi level along the $k_{x}$ and $k_{y}$ axes are purely from the $d_{xz}$ orbitals, as demonstrated by the matrix elements analysis (Fig. S2) and the band structure calculations (Fig. 3c and 3d). We note that there is a slight band hybridization of $d_{xz}$ and $d_{yz}$ orbitals on some portions of the Fermi surface with an Fermi surface angle between $\theta$=15$^\circ$$\sim$45$^\circ$ (Fig. 3c and 3d). But even in this case, the $d_{xz}$ orbital still dominates the electronic states at the Fermi level. There have been a controversy regarding the orbital nature of the electronic states near the Fermi level\cite{Watson_PRB,watson2_PRB,Suzuki_PRB,Sprau_arxiv}. Our results demonstrate unambiguously that it is the $d_{xz}$ orbital that dominates the low energy electronic structure of FeSe around $\Gamma$ point in the nematic state.

%Figure4
After resolving the dominant $d_{xz}$ orbital nature that dictates the Fermi surface and superconducting gap of FeSe superconductor, now we come to investigate some key factors that are closely related to its superconductivity. Fig. 4a shows the momentum distribution curves (MDCs) at the Fermi level along various Fermi surface angles $\theta$ (schematically shown in top-left inset in Fig. 4a) measured in normal state at 9.8 K. The spectral weight (red region in Fig. 4a) is strongest near the $\theta$=90$^\circ$ and gets weaker when moving to $\theta$=0$^\circ$ and 180$^\circ$. Quantitative analysis of the spectral weight (MDC area) as a function of the Fermi surface angle is plotted in Fig. 4d; it is peaked near $\theta$=90$^\circ$ and 270$^\circ$. Fig. 4b shows photoemission spectra (EDCs) along the Fermi surface measured at 9.8 K in normal state and the spectral weight of the $d_{xz}$ orbital (red region in Fig. 4b) at different Fermi surface angles is plotted in Fig. 4e. It shows similar variation with the MDC weight. The analysis in the superconducting state gives similar results (Fig. S4). For comparison, the superconducting gap of FeSe as a function of the Fermi surface angle measured at 1.6 K is re-plotted in Fig. 4c. Overall the superconducting gap exhibits an anti-correlation relation with the spectral weight of the $d_{xz}$ orbital, i.e., the gap minimum corresponds to the spectral weight maximum. We also analyzed the effective mass of the $\alpha$ band as the function of the Fermi surface angle, as shown in Fig. 4f. The effective mass displays a dramatic variation along the Fermi surface; its value along the $k_{x}$ axis is nearly ten times that along the $k_{y}$ axis. It also shows a perfect anti-correlation with the superconducting gap: the gap maximum (minimum) corresponds to the minimum (maximum) of the effective mass.

%Contrast to BQPI paper on the spectral weight

%Discussion the origin of anisotropic gap
The present observations provide key information on understanding the relationship between nematicity, electronic structure and superconductivity in FeSe. In the nematic state, the splitting of the $d_{xz}$ and $d_{yz}$ orbitals gives rise to an anisotropic Fermi surface around the $\Gamma$ and X/Y points (corresponding to BZ of 1-Fe unit cell)\cite{Watson_PRB,watson2_PRB}. In the extreme case when the low energy electronic states are dominated by a single $d_{xz}$ orbital, the electron hopping along the $x$ direction is much more enhanced than that along the $y$ direction (Fig. 4h), resulting in highly anisotropic Fermi surface topology (Fig. 1c) and effective mass (Fig. 4f) in the nematic state of FeSe. Since nearly isotropic superconducting gap with four-fold symmetry is observed in tetragonal Fe(Se,Te) superconductors\cite{Okazaki_PRL,HMiao_PRB}, the observed extremely anisotropic superconducting gap with two-fold symmetry in FeSe is expected to be associated with its nematicity. The question comes to how nematicity can generate such an anisotropic superconducting gap. One scenario proposed involves the splitting of the $d_{xz}$ and $d_{yz}$ orbitals\cite{JKang_PRL,Agatsuma_PRB}. The orbital order in the nematic state introduces a $d$-wave component on top of the existing $s$-wave component, leading to a highly anisotropic superconducting gap with even accidental nodes\cite{JKang_PRL}. Our measured superconducting gap (Fig. 2d) is consistent with the expected $es$+$d$ type gap form in this picture\cite{JKang_PRL}. However, it is expected that the superconducting gap is positively correlated with the orbital occupation\cite{Agatsuma_PRB}, which is opposite to our observation that the gap and the spectral weight exhibit an anti-correlation (Fig. 4c-4e). This picture involves two hole-pockets around the $\Gamma$ point, each of which is composed of both $d_{xz}$ and $d_{yz}$ orbitals\cite{JKang_PRL,Agatsuma_PRB}. It needs to be modified in order to make a direct comparison with our results of FeSe where there is only one Fermi pocket around the $\Gamma$ point which is dominated by a single $d_{xz}$ orbital.

To understand the anisotropic superconducting gap of FeSe, another scenario proposed considers orbital-selective Cooper pairing in a modified spin-fluctuation theory via suppression of pair scattering processes involving those less coherent states\cite{Sprau_arxiv}. This picture can produce a superconducting gap structure that is consistent with our results (Fig. 2d) and previous experimental observations\cite{Sprau_arxiv,Xu_PRL}. It was proposed that the pairing is mainly based on electrons from the $d_{yz}$ orbital of the iron atoms\cite{Sprau_arxiv,Kreisel_PRB}. Our observations provide direct evidence that, in the nematic state of FeSe, the extremely anisotropic superconducting gap opens on the highly anisotropic Fermi surface (Fig. 4g) that is dominated by the $d_{xz}$ orbital. In FeSe, it was also proposed that the superconducting gap shows a positive correlation with the spectral weight of the $d_{yz}$ orbital that is considered to be responsible for superconductivity\cite{Sprau_arxiv}. From our direct ARPES results, although we find that the spectral weight of the $d_{yz}$ orbital exhibits a good correspondence to the superconducting gap (Fig. S5), the $\beta$ band with $d_{yz}$ orbital character lies $\sim$20 meV below the Fermi level, thus contributing little to superconductivity of FeSe. Our results indicate that superconductivity in FeSe is dictated by the $d_{xz}$ orbital and its spectral weight shows an anti-correlation with the superconducting gap: the gap minimum corresponds to the spectral weight maximum of the $d_{xz}$ orbital (Fig. 4c-4e). Our results ask for a reexamination on the picture of the orbital-selective superconductivity in FeSe\cite{Sprau_arxiv}.

The fact that the measured gap function can be fitted to a simple $p$-wave gap function also leads us to ask whether the spin-triplet pairing is possible in FeSe. In the nematic state, because of the dominant single $d_{xz}$ orbital feature of the Fermi surface around the $\Gamma$ and the strong Hund's rule coupling, the interaction of the $d_{xz}$ band with other bands near X/Y is expected to become rather weak. For the $d_{xz}$ orbital, the dominant intra-orbital coupling is the nearest neighbor magnetic exchange couplings along the $x$ axis, namely the $J_{1x}$-interaction shown in Fig. 4h, which is ferromagnetic in iron-chalcogenides from neutron scattering measurements\cite{Johnston2010,Hu2012,Wang2012,Dai2015}. If this is the strongest pairing interaction, the superconducting pairing on the Fermi surface is possible to have a triplet $p$-wave symmetry with a gap function proportional to $cos(\theta)$, which is consistent with our observations. No drop of the Knight shift across T$_c$ is observed in NMR measurements of bulk FeSe, which is also compatible with a possible triplet pairing\cite{AEBohmer_PRL,Baek_NM}. In addition to the coexisting N\'{e}el and stripe spin fluctuations observed by neutron scattering\cite{QWang_NM,QWang_NC}, recent observation of charge ordering in FeSe suggests a presence of an additional magnetic fluctuation with a rather small wavevector\cite{WLi_NP} that is related to intra-pocket scattering around $\Gamma$ point\cite{YTTam}. Our present work raises an interesting possibility of $p$-wave pairing symmetry in FeSe. We note that ARPES can only measure the magnitude of the superconducting order parameter. Also because of the limitation of the laser photon energy (6.994 eV), it is not possible for us to measure the superconducting gap on the Fermi surface around the X/Y point (Fig. 1b). Considering that the measured superconducting gap can be fitted quite well by both a simple $p$-wave form (green line in Fig. 2d) and a combined gap form of an anisotropic $s$- and $d$-waves (red line in Fig. 2d), it is clear that further experimental and theoretical studies are needed to establish the pairing symmetry of FeSe.

In summary, by carrying out high resolution laser-based ARPES on the bulk FeSe, we have observed an extremely anisotropic superconducting gap on the highly anisotropic Fermi surface in the nematic state of FeSe superconductor. The Fermi pocket around $\Gamma$ is dominated by $d_{xz}$ orbital. The spectral weight of the $d_{xz}$ orbital shows an anti-correlation with the superconducting gap along the Fermi surface. Our results directly demonstrate that the Cooper pairing on the Fermi pocket around $\Gamma$ originates from electrons of $d_{xz}$ orbitals in the nematic state of FeSe. They provide key insights on the pairing symmetry and the interplay between nematicity and superconductivity in iron-based superconductors.

{\bf Methods}

High quality FeSe bulk single crystals were grown by KCl/AlCl$_{3}$ chemical vapor transport technique\cite{BLei_PRL,Bohmer_PRB}. The samples are characterized by x-ray diffraction (XRD) (Fig. S1a). Electrical resistance measurement and magnetic measurement (Fig. S1b, c) show a T$_{c}$ of 8.0 K with a sharp transition width of $\sim$0.4 K.

High-resolution ARPES measurements were performed at our new laser-based system equipped with the 6.994 eV vacuum ultraviolet laser and the time-of-flight electron energy analyser (ARToF 10k by Scienta Omicron)\cite{YZhang_NC}. This latest-generation ARPES system is capable of measuring photoelectrons covering two-dimensional momentum space ($k_x$, $k_y$) simultaneously. The system is equipped with an ultra-low temperature cryostat which can cool the sample to a low temperature of 1.6 K. Measurements were performed using both $LH$ and $LV$ polarization geometries (see Fig. S2). The energy resolution is $\sim$1 meV and the angular resolution is $\sim$0.1 degree. All the samples were measured in ultrahigh vacuum with a base pressure better than 5.0$\times$10$^{-11}$ mbar. The samples were cleaved {\it in situ} and measured at different temperatures. The Fermi level is referenced by measuring polycrystalline gold which is in good electrical contact with the sample, as well as the normal state measurement of the sample above T$_{c}$.

To simulate the band structure of FeSe, we adopted 5-orbital tight-binding model including onsite spin-orbital coupling ($\lambda$). In the nematic state, we further consider $s$-wave ($\Delta_{s}$) and $d$-wave ($\Delta_{d}$) orbital order which break $C_4$ rotational symmetry. The Hamiltonian of these two orders are given by,
\begin{equation}
%%H_{OO}=\sum_{\boldsymbol{k}}\Delta_{s}(cosk_{x}+cosk_{y})(n_{xz}(\boldsymbol{k})-n_{yz}(\boldsymbol{k}))+\sum_{\boldsymbol{k}}\Delta_{d}(cosk_{x}-cosk_{y})(n_{xz}(\boldsymbol{k})+n_{yz}(\boldsymbol{k}))
H_{s}=\sum_{\boldsymbol{k}}\Delta_{s}(cosk_{x}+cosk_{y})(n_{xz,\boldsymbol{k}}-n_{yz,\boldsymbol{k}})
\end{equation}
\begin{equation}
%%H_{OO}=\sum_{\boldsymbol{k}}\Delta_{s}(cosk_{x}+cosk_{y})(n_{xz}(\boldsymbol{k})-n_{yz}(\boldsymbol{k}))+\sum_{\boldsymbol{k}}\Delta_{d}(cosk_{x}-cosk_{y})(n_{xz}(\boldsymbol{k})+n_{yz}(\boldsymbol{k}))
H_{d}=\sum_{\boldsymbol{k}}\Delta_{d}(cosk_{x}-cosk_{y})(n_{xz,\boldsymbol{k}}+n_{yz,\boldsymbol{k}})
\end{equation}
where $n_{\alpha,\boldsymbol{k}}=n_{\alpha,\boldsymbol{k}\uparrow}+n_{\alpha,\boldsymbol{k}\downarrow}$ is the density for $\alpha$ orbital. To match the data in ARPES experiments in the nematic state, the adopted parameters in the calculations are $\lambda$=10 meV, $\Delta_{s}$=17.5 meV and $\Delta_{d}$=-10 meV. In normal state, we set $\lambda$=10 meV, $\Delta_{s}$=0 meV and $\Delta_{d}$=0 meV.\\

\vspace{3mm}

\noindent {\bf Acknowledgement}\\
We thank Dunghai Lee, Rafael Fernandes, Chandra Varma, Qimiao Si, Guangming Zhang, J. C. Seamus Davis, Zhixun Shen and Guoqing Zheng for helpful discussions. This work is supported by the National Key Research and Development Program of China (Grant No. 2016YFA0300300 and 2017YFA0302900), the National Natural Science Foundation of China (Grant No. 11334010 and 11534007), the National Basic Research Program of China (Grant No. 2015CB921000 and 2015CB921300) and the Strategic Priority Research Program (B) of the Chinese Academy of Sciences (Grant No. XDB07020300 and XDPB01).

\vspace{3mm}

\noindent {\bf Author Contributions}\\
D.F.L., C.L., J.W.H. contributed equally to this work. X.J.Z. and D.F.L. proposed and designed the research. B.L., L.W., Y.G.S. and X.H.C. contributed to FeSe crystal growth. X.L.D contributed to the magnetic measurement. X.X.W., J.P.H. and T.X. contributed to the band structure calculations and theoretical discussion. D.F.L., C.L., J.W.H., B.S., Q.G., Y.X.Z., X.L., Y.H., Y.X., A.J.L., J.L., P.A., L.Z., S.L.H., L.Y., G.D.L., F.F.Z., S.J.Z., F.Y., Z.M.W., Q.J.P., Z.Y.X., C.T.C. and X.J.Z. contributed to the development and maintenance of Laser-ARTOF system. X.W.J. contributed to software development for data analysis. D.F.L., C.L., J.W.H. and B.S. carried out the experiment with the assistance from Q.G., Y.X.Z., X.L., Y.H., Y.X., A.J.L., J.L. and P.A.. D.F.L., C.L., J.W.H. and X.J.Z. analyzed the data. D.F.L. and X.J.Z. wrote the paper with C.L., J.W.H., J.P.H. and T.X.. All authors participated in discussion and comment on the paper.

%%\vspace{3mm}

%%\noindent {\bf\large Additional information}\\
%%\noindent{\bf Competing financial interests:} The authors declare no competing financial interests.

\newpage

\begin{figure*}[tbp]
\begin{center}
\includegraphics[width=1.0\columnwidth,angle=0]{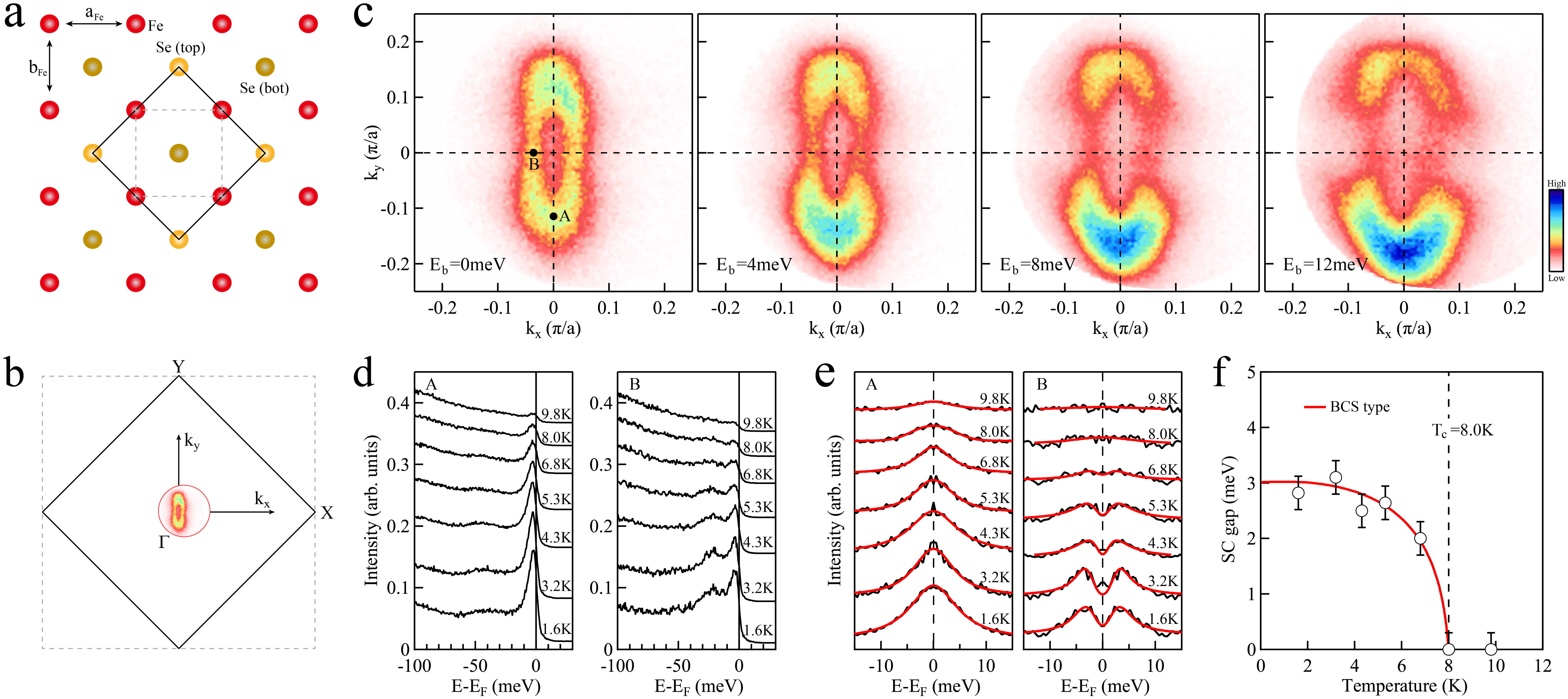}
\end{center}
\caption{\textbf{Fermi surface of the single domain FeSe measured at 1.6 K around the \boldsymbol{$\Gamma$} point and its superconducting gap.} (a) Top view of FeSe crystal structure. The red, yellow and gray circles represent Fe atom, Se atom above the Fe plane (Se top) and Se atom below the plane (Se bottom), respectively. The dashed line represents the 1-Fe unit cell while the solid line represents 2-Fe unit cell. We define $x$ axis parallel to $a_{Fe}$ axis and $y$ axis parallel to $b_{Fe}$ axis. In the normal state, FeSe has $C_{4}$ symmetry with $a_{Fe}$=$b_{Fe}$, but it breaks into $C_{2}$ symmetry in the nematic state with $a_{Fe}$$>$$b_{Fe}$. (b) The Brillouin zone of FeSe. The dashed line and the solid line correspond to BZs of the 1-Fe unit cell and 2-Fe unit cell, respectively. The red circle represents the momentum area that can be measured simultaneously by our laser-ARPES system with 6.994 eV photon energy. The whole Fermi surface of FeSe around the $\Gamma$ point can be covered by one time measurement which occupies a very small portion of the BZ. (c) Constant energy contours around the $\Gamma$ point at different binding energies measured at 1.6 K in $LV$ polarization geometry. The measured Fermi surface (leftmost panel) shows an elliptical shape with its long axis along $k_{y}$ axis and short axis along $k_{x}$ axis. The spectral weight concentrates to the vertex areas and gets depleted around the central area with increasing binding energy. (d) Photoemission spectra (EDCs) measured at different temperatures for two typical Fermi momenta. The location of the two momenta, A and B, is marked in (c) as black dots. (e) Corresponding symmetrized EDCs from (d). The curves are fitted with a phenomenological gap formula\cite{Norman_PRB} (solid red curves). For point A (left panel), no gap opening is detected below T$_{c}$ within our experimental resolution. Apparent gap opening occurs for point B (right panel). (f) Temperature dependence of the superconducting gap at B point. It follows the BCS gap form (red curve).
}
\end{figure*}

\begin{figure*}[tbp]
\begin{center}
\includegraphics[width=0.85\columnwidth,angle=0]{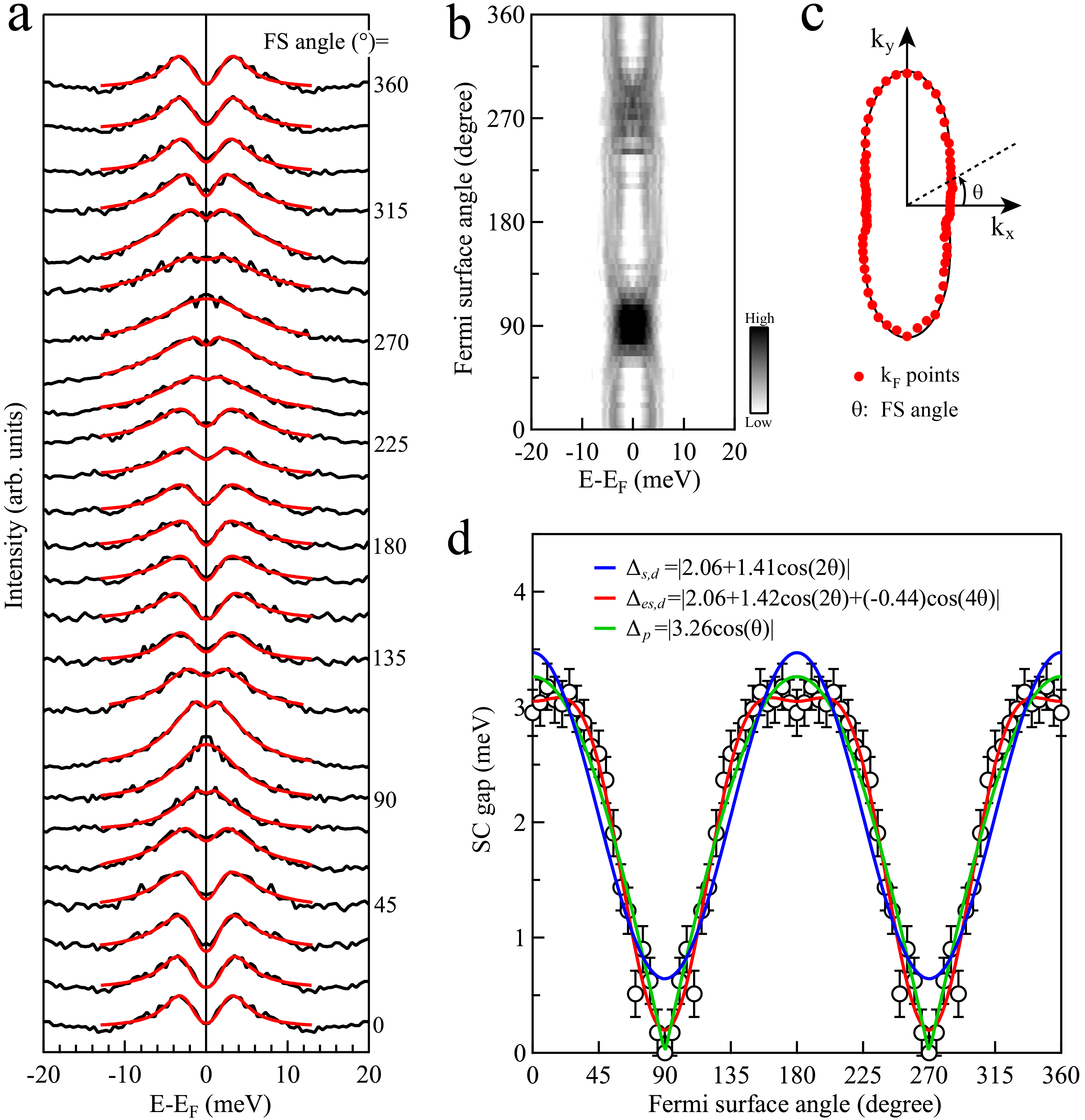}
\end{center}
\caption{\textbf{Momentum dependence of the superconducting gap for FeSe measured at 1.6 K.} (a) Symmetrized EDCs at different Fermi momenta along the Fermi surface measured at 1.6 K. These curves are fitted with a phenomenological gap formula (red curves). The same data are plotted in (b) as a false color image where the variation of the superconducting peak position can be directly visualized. The location of the Fermi momentum is defined by the Fermi surface angle $\theta$ as shown in (c). (d) Momentum dependence of the superconducting gap derived from fitting the symmetrized EDCs in (a). To enhance the data statistics and keep the two-fold symmetry, the gap value is obtained by averaging over the four quadrants. The measured gap (empty circles) is fitted by $s$+$d$ (blue curve), extended-$s$+$d$ (red curve) and $p$ (green curve) wave pairing gap forms.
}
\end{figure*}

\begin{figure*}[tbp]
\begin{center}
\includegraphics[width=1.0\columnwidth,angle=0]{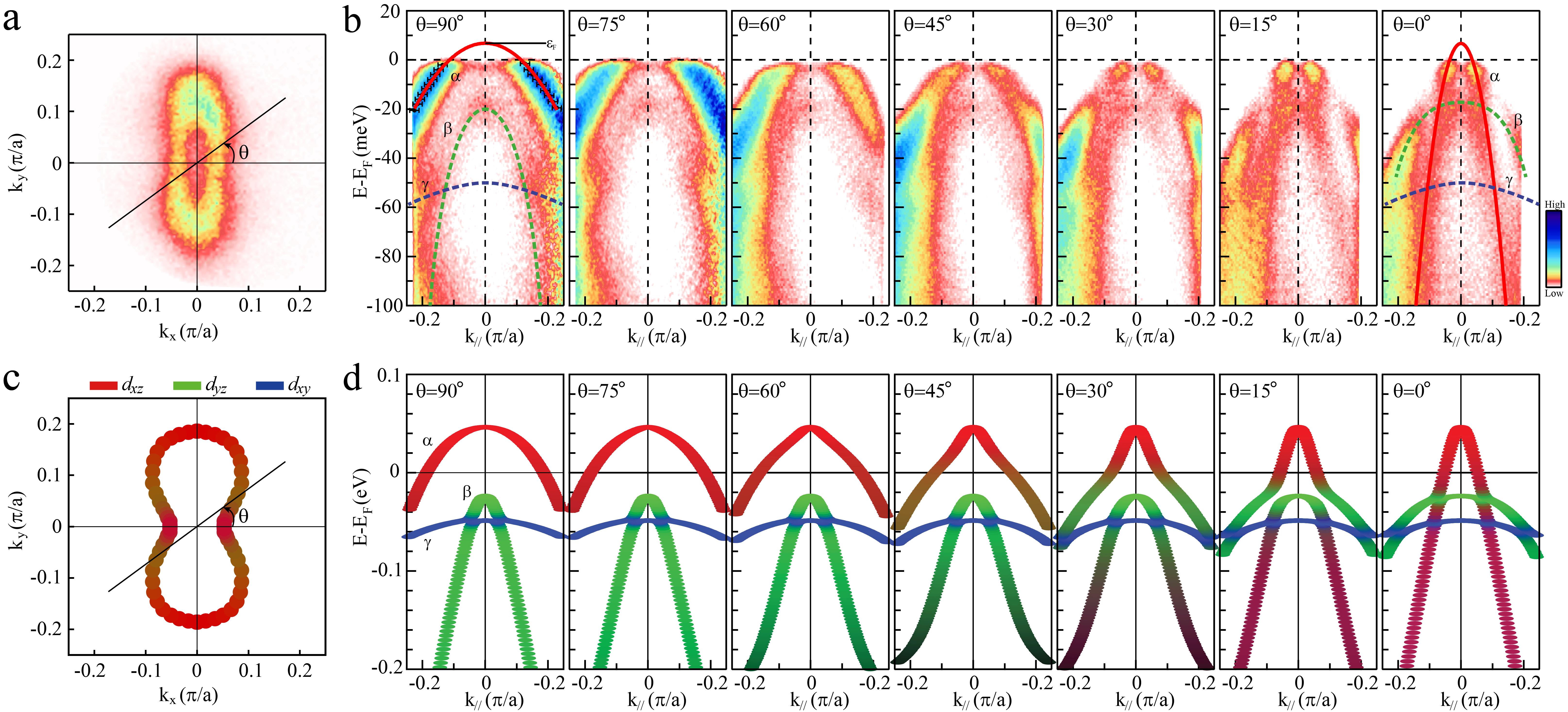}
\end{center}
\caption{\textbf{Band structure of FeSe along different momentum cuts and their orbital nature.} (a) Definition of the momentum cuts across the $\Gamma$ point by the Fermi surface angle $\theta$. (b) Measured band structure along different momentum cuts varying from $\theta$=90$^\circ$ (vertical cut, leftmost panel) to $\theta$=0$^\circ$ (horizontal cut, rightmost panel). Three bands are observed, marked as $\alpha$, $\beta$ and $\gamma$ in the $\theta$=90$^\circ$ and $\theta$=0$^\circ$ panels. The $\alpha$ band is well fitted by a parabolic curve (red line) which gives a Fermi energy $\varepsilon_{F}$ of $\sim$6.7 meV above the Fermi level. The effective mass for different momentum cuts can be obtained from such a parabolic fitting. The dashed green and blue lines are guides to the eye for the $\beta$ and $\gamma$ bands, respectively. (c) The calculated Fermi surface of FeSe around $\Gamma$ point in the nematic state. (d) The calculated band structure along different momentum cuts defined by the Fermi surface angle $\theta$ in (c). In (c) and (d), the $d_{xz}$, $d_{yz}$ and $d_{xy}$ orbital components are represented by red, green and blue colors, respectively.
}
\end{figure*}

\begin{figure*}[tbp]
\begin{center}
\includegraphics[width=0.8\columnwidth,angle=0]{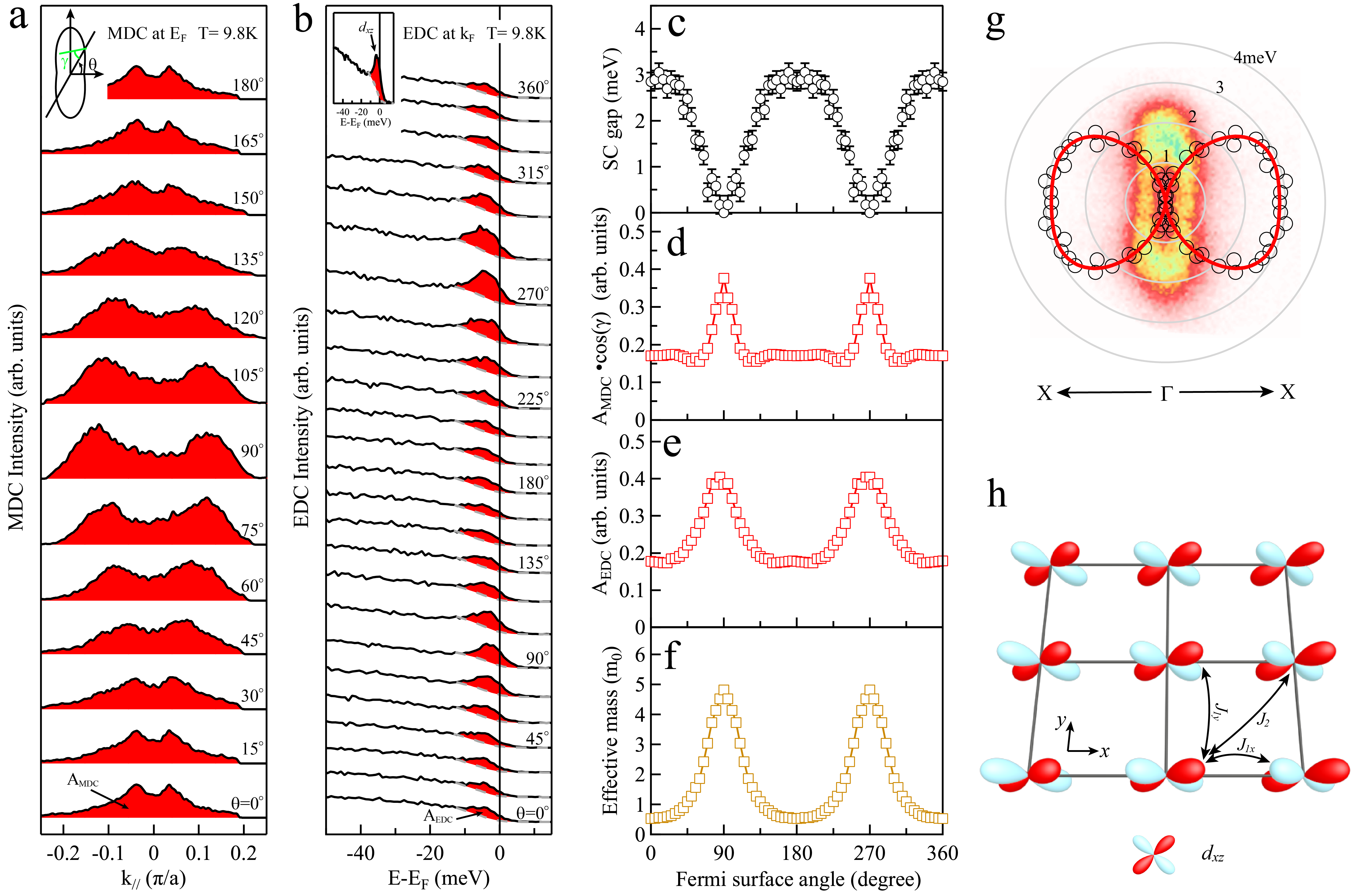}
\end{center}
\caption{\textbf{Correlation between the superconducting gap, the spectral weight and the effective mass of bands along the Fermi surface of FeSe.} (a) Momentum distribution curves (MDCs) along different momentum cuts at the Fermi level measured at 9.8 K. The momentum cut is defined by the Fermi surface angle $\theta$ as shown in the inset. For a given Fermi momentum on the Fermi surface, the angle between the momentum cut and the Fermi surface normal direction is defined as an angle $\gamma$. The area of the MDCs marked in red is defined as the spectral weight $A_{MDC}$ in (d). (b) Photoemission spectra (EDCs) of FeSe along the Fermi surface measured at 9.8 K. After subtracting the Shirley background (dashed line), the area of the EDCs marked in red is defined as the spectral weight $A_{EDC}$ in (e). For the convenience of comparison, the superconducting gap measured at 1.6 K is re-plotted in (c). (d) MDC spectral weight along the Fermi surface obtained from (a) after considering the correction of the $\gamma$ angle as defined in inset of (a). (e) EDC spectral weight along the Fermi surface obtained from (b). (f) Effective mass of the $d_{xz}$ band along different momentum cuts as obtained by a parabolic fitting of the $\alpha$ band in Fig. 3b. (g) Superconducting gap plotted in the polar graph, overlaid with the measured Fermi surface in order to have a direct comparison of their relative orientations. (h) A schematic picture of the $d_{xz}$ orbitals in the Fe plane in the nematic state of FeSe. For the $d_{xz}$ orbitals, the dominant intra-orbital coupling is the exchange interactions along the $x$ axis ($J_{1x}$). Both $J_{1y}$ and $J_2$ are antiferromagnetic-like while the $J_{1x}$ coupling is ferromagnetic in iron-chalcogenides\cite{Johnston2010,Hu2012,Wang2012,Dai2015}.
}
\end{figure*}

\end{document}